\documentstyle [twocolumn,epsf]{mn}
\newcommand{\mincir}{\raise
-2.truept\hbox{\rlap{\hbox{$\sim$}}\raise5.truept\hbox{$<$}\ }}
\newcommand{\magcir}{\raise
-2.truept\hbox{\rlap{\hbox{$\sim$}}\raise5.truept\hbox{$>$}\ }}
\newcommand{\minmag}{\raise
-2.truept\hbox{\rlap{\hbox{$<$}}\raise6.truept\hbox{$<$}\ }}
\newcommand{\be}{\begin{equation}}
\newcommand{\ee}{\end{equation}}
\newcommand{\ba}{\begin{eqnarray}}
\newcommand{\ea}{\end{eqnarray}}
\newcommand{\brr}{\begin{array}}
\newcommand{\err}{\end{array}}
\newcommand{\bc}{\begin{center}}
\newcommand{\ec}{\end{center}}

\title[Large-Scale Anisotropy]{Large-Scale Coherent Dipole Anisotropy ?}
\author[S.Basilakos \& M.Plionis]
{S.Basilakos$^{1,2}$ \& M. Plionis$^{1}$ \\
$^1$ National Observatory of Athens, Lofos Nimfon, Thesio, 18110 Athens,
Greece \\
$^2$ Physics Dept., University of Athens, Panepistimiopolis, Greece \\
}

\begin{document}
\maketitle

\begin{abstract}
We have reanalyzed and compared the dipoles of the 1.2 Jy and 0.6 Jy (QDOT) 
IRAS  galaxy samples. We find strong indications from both samples 
for (a) significant contributions to the gravitational field that shapes
the Local Group motion from depths up to $\sim 170$ $h^{-1}$ Mpc and (b) 
a large-scale coherence of the dipole anisotropy, indications provided mainly
by the fact that the differential dipoles of large equal volume shells are  
aligned with the CMB dipole and exhibit significant dipole signals.
The two IRAS dipoles are indistinguishable within 50 $h^{-1}$ Mpc and 
beyond $\sim 130$ $h^{-1}$ Mpc while the QDOT dipole, having a lower flux 
limit, continues growing with respect to the 1.2 Jy sample up to $\sim 100$ 
$h^{-1}$ Mpc in agreement with Rowan-Robinson et al (1990). 

\vspace{0.35cm}

\end{abstract}

\section{Introduction}

The peculiar velocity of the Local Group of galaxies with respect to the 
Cosmic Microwave Background (CMB) with $u_{LG}=$622 km/sec 
towards $(l,b)=(277^{\circ},30^{\circ})$ is a well established fact 
(cf. Kogut et al. 1993). 
The most probable cause for this motion as 
well as for the observed peculiar motions of other galaxies and clusters (cf. 
Dekel 1997 and references therein) is gravitational instability (cf. Peebles 
1980).
This is supported by the fact that the gravitational dipole (acceleration) of
many different samples of extra-galactic mass tracers is well aligned with
the general direction of the CMB dipole (cf. Yahil, Walker \& 
Rowan-Robinson 1986; Lahav 1987; Lynden-Bell et al 1988; Miyaji \& Boldt 1990;
Rowan-Robinson et al 1990; Strauss et al 1992; Hudson 1993; 
Scaramella et al 1991; Plionis \& Valdarnini 1991; Branchini \& Plionis 1996).
However, what still
seems to be under discussion is from which depths do density fluctuations 
contribute to the gravitational field that shapes the Local Group motion.
The largest such depth is defined by the dipole {\em convergence 
depth}, $R_{conv}$, which is that depth where the true gravitational 
acceleration converges to its final value. The outcome of many studies, using 
different 
flux or magnitude limited galaxy samples, is that the apparent value of
$R_{conv}$ differs from 
sample to sample, in the range from 40 to 100 $h^{-1}$ Mpc, with a strong 
dependence to the sample's characteristic depth. This probably implies that 
the apparent dipole convergence is spurious, due to lack of adequately 
sampling the distant density fluctuations. 
Only the optical Abell/ACO cluster sample is volume limited out to a large 
enough depth ($\simeq 240 \; h^{-1}$ Mpc) to allow a more reliable 
determination of $R_{conv}$ which was found to be $\simeq 160\; h^{-1}$ Mpc
(Scaramella et al 1991; Plionis \& Valdarnini 1991; Branchini \& Plionis 
1996). Recently, this result has been 
confirmed using X-ray cluster samples, which are free of the various 
systematic effects from which the optical catalogues suffer (Plionis \& 
Kolokotronis 1998).

If there is a linear bias relation between the cluster, the galaxy and the 
underlying matter density fluctuations, as usually assumed (cf. Kaiser 1984), 
then the galaxy dipole should also have similarly deep contributions. 
In this study we reanalyse the 1.2 Jy and the deeper QDOT 0.6 Jy IRAS galaxy 
dipoles, initially investigated by Strauss et al (1992) and Rowan-Robinson et 
al (1990) respectively, with the aim of investigating whether 
there are any such indications.

\section{IRAS galaxy samples and selection functions}
We use in our analysis the two available flux-limited 60-$\mu$m IRAS samples; 
one limited at $S_{lim}=1.2$ Jy (Fisher et al 1995) and the
other at $S_{lim}=0.6$ Jy (Rowan-Robinson et al 1990), which has a 1 in 6 
sampling rate.
The IRAS 1.2 Jy contains 5763 galaxies with $|b|>5^{\circ}$ while the QDOT 
contains 2086 galaxies with $|b|>10^{\circ}$. Note that although the two 
catalogues are not totally independent, a cross correlation revealed only 
105 common galaxies (with $\delta\theta \le 0.6^{\circ}$ and 
$\delta cz \le 800$ km/sec). 

To estimate the local acceleration field it is necessary to recover the true 
galaxy density field from the observed flux-limited samples. This is done
by weighting each galaxy by $\phi^{-1}(r)$, where the selection
function, $\phi(r)$, is defined as the fraction of the galaxy number density 
that is observed above the flux limit at some distance $r$. Therefore
\begin{equation}\label{eq:sf}
\phi(r)=\frac{1}{\langle n_{g} \rangle} \int_{L_{min}(r)}^{L_{max}} \Phi(L) dL
\end{equation} 
where $L_{min}(r)=4\pi r^{2} \nu S_{lim}$ is the luminosity of a source 
at distance $r$ corresponding to the flux limit $S_{lim}$, $\nu =$ 60-$\mu$m 
and $\langle n_{g} \rangle$ is the mean galaxy number density, given by 
integrating the luminosity function over the whole luminosity range, with
$L_{min}=7.5\times 10^{7} \; h^{2} L_{\odot}$ since lower luminosity galaxies
are not represented well in the available samples (cf. Rowan-Robinson et al 
1990; Fisher et al 1995), and $L_{max}= 10^{13} \; 
h^{2} L_{\odot}$. Obviously, $\phi(r)$ is a 
decreasing function of distance because a smaller 
fraction of the luminosity function falls above the flux limit at greater 
distances. 

For the QDOT sample we used the Saunders et al (1990) luminosity function
while for the IRAS 1.2 Jy we used the parameterised selection function of
Yahil et al (1991). We have verified, however, that the two selection functions
are indistinguishable from each other when applied to the same flux limit.
\begin{figure}
\mbox{\epsfxsize=8cm \epsfysize=7cm \epsffile{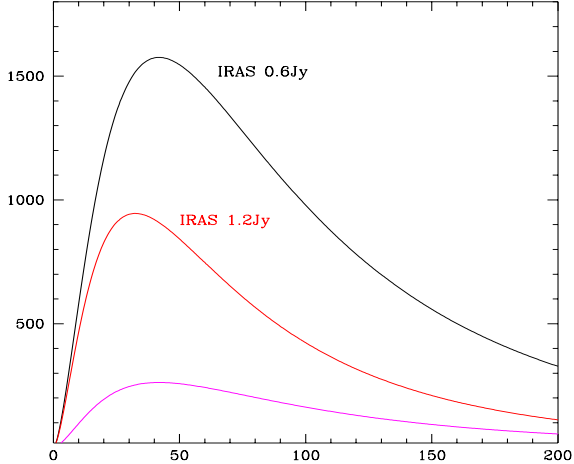}}
\caption{The expected $N(r)$ distribution, according to the IRAS galaxy 
luminosity function, for the 1.2 Jy, the QDOT 0.6 Jy and the PSCz 0.6 Jy IRAS 
samples.}
\end{figure}
In figure 1 we present the $\phi(r)$ of the IRAS 1.2 and 0.6 Jy samples,
for the 1-in-6 (QDOT) as well as for the unavailable 6-in-6 sampling 
rate (PSCz). 
It is evident that although the QDOT sample is deeper, it samples the galaxy 
distribution more sparsely than the 1.2 Jy sample.

\section{Dipole calculation} 
We determine the peculiar acceleration of Local Group galaxies by 
measuring moments of the IRAS galaxy distribution.
The dipole moment:
${\bf D}=\sum \phi^{-1}(r) r^{-2} \hat{r}$,
is calculated by weighing
the unit directional vector pointing to the position of each galaxy, with the 
gravitational weight of that galaxy and summing over all available galaxies
with distances greater than 5 $h^{-1}$Mpc (on smaller scales the 
observed galaxies do not adequately represent the true distribution; cf. 
Rowan-Robinson et al 1990). Similarly the monopole term is:
$M=\sum \phi^{-1}(r) r^{-2}$.

We then estimate the gravitational acceleration induced on the LG from
the distribution of IRAS galaxies by:
\begin{equation}\label{eq:dip}
{\bf V}_{g}(r)=\frac{H_{\circ} R_{conv}}{M(\le R_{conv})} 
{\bf D}(r)
\end{equation}
(cf. Miyaji \& Boldt 1990; Plionis et al 1993).
Using linear perturbation theory (cf. Peebles 1980) and equation 
(\ref{eq:dip}) we can relate the Local Group peculiar velocity with the 
estimated acceleration by: 
\begin{equation}\label{eq:lpt}
{\bf u}_{\footnotesize LG}(r)=\beta_{I}{\bf V}_{g}(r)
\end{equation} 
where $\beta_{I}=\Omega^{0.6}/b_{I}$ and $b_{I}$ is the IRAS
galaxy to underlying mass bias factor.
 
\subsection{Treatment of the IRAS galaxy data}
Due to systematic effects and biases present in the data we have to 
perform various corrections to the raw dipole estimates.
Firstly we need to treat the excluded, due to cirrus emission, galactic plane.
We do so by extrapolating to these regions the data from the rest of the unit 
sphere with the help of a spherical harmonic expansion of the galaxy surface 
density field and a sharp mask (cf. Yahil et al 1986; Lahav 1987).
Secondly, about $4\%$ of the sky is not covered by the catalogue and we apply 
to these areas a homogeneous distribution of galaxies having the mean 
weight, estimated from the rest of the sky. 
Thirdly, due to discreteness 
effects and the steep selection function with depth we have an additive 
dipole term, the {\em shot-noise dipole}, for which we have to correct our 
raw dipole estimates. Assuming Gaussianity, the Cartesian 
components of the shot noise dipole are equal $(\sigma_x= \sigma_y=
\sigma_z$) and thus $\sigma_{3D}^{2}= 3 \sigma_{i,1D}^{2}$
(cf. Hudson 1993). 
Taking the coordinate system such that one of the shot-noise dipole components
is parallel to the z-axis of the true dipole and we can attempt an approximate
correction of the raw dipole according to the following model:  
\begin{equation}
D_{cor}=D_{raw}-\sigma_{3D}/\sqrt{3}
\end{equation}
Note that this correction model although more severe than the usual 
$D_{cor}^{2}=D_{raw}^{2}-\sigma_{3D}^{2}$ model, it provides qualitatively 
similar dipole corrections. We choose, however, to use this model in order to 
be conservative and to obtain a sort of lower limit to the resulting 
dipole, as far as the shot-noise correction is concerned, and thus via 
eq.(\ref{eq:lpt}) an upper limit to the estimated cosmological $\beta$ 
parameter (see section 4.4).

To calculate $\sigma_{3D}$ we use two methods; a Monte-Carlo simulation 
approach in which we randomise the angular coordinates of all galaxies while 
keeping their distance, and thus their selection function, unchanged 
while the second method is the analytic estimation of Strauss et al (1992):  
$\sigma_{3D}^{2} \simeq \sum \phi_{i}^{-1}r_{i}^{-4} (\phi_{i}^{-1}+1)$.
\begin{figure}
\mbox{\epsfxsize=8cm \epsfysize=7cm \epsffile{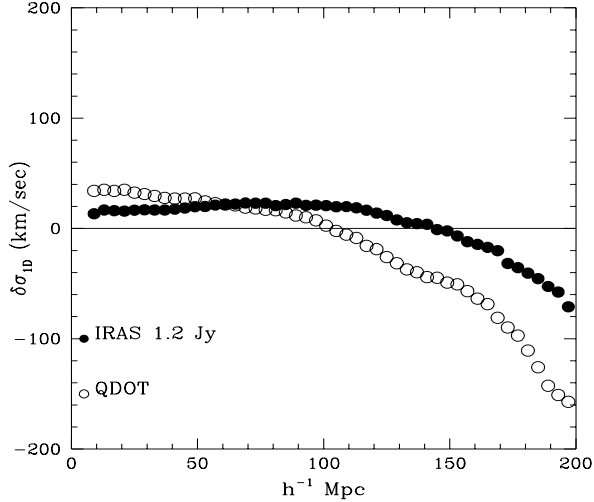}}
\caption{Difference between the Monte-Carlo and the Strauss et al (1992) 
1-d shot-noise estimates, for both IRAS 1.2 Jy and QDOT samples.}
\end{figure}
Figure 2 shows the difference (in velocity units) between the two shot-noise 
estimates. It is evident that the two methods give equivalent results 
although due to the statistical nature of the first method we believe that it 
performs better on large depths, where the number density of IRAS galaxies is 
very low. 

\subsection{$z$ to $3d$ frame correction}
The final but essential correction is to transform redshifts to 3-d distances
in order to minimise the so called `Kaiser' effect (Kaiser 1987).
This effect can be understood by noting that the distribution
of galaxies in redshift space differs from that in real comoving
space by a non-linear term:
\begin{equation}\label{eq:hubble}
cz=H_{\circ}r+({\bf v}(r)-{\bf v}(0)) \cdot \hat{r}
\end{equation}
where ${\bf v}(0) (\equiv {\bf u}_{LG})$ is the peculiar velocity of
the Local Group and ${\bf v}(r)$ the peculiar velocity of a galaxy at position
${\bf r}$. If ${\bf v}(r)$ had random orientation, then 
$\int {\bf v}(r) \cdot \hat{\bf r} \; \rm{d}^{3}r \approx 0$ 
and the last term of eq.(\ref{eq:hubble}) is dominated by the LG term; we 
thus obtain that in the LG frame (ie., when using $cz=H_{\circ} r$), 
structures in the direction of 
our motion appear at a redshift smaller than their true distance in the CMB 
frame and thus they will artificially enhance the amplitude of the 
gravitational dipole. 
\begin{figure}
\mbox{ \epsfysize=12cm \epsfxsize=10cm 
\epsffile{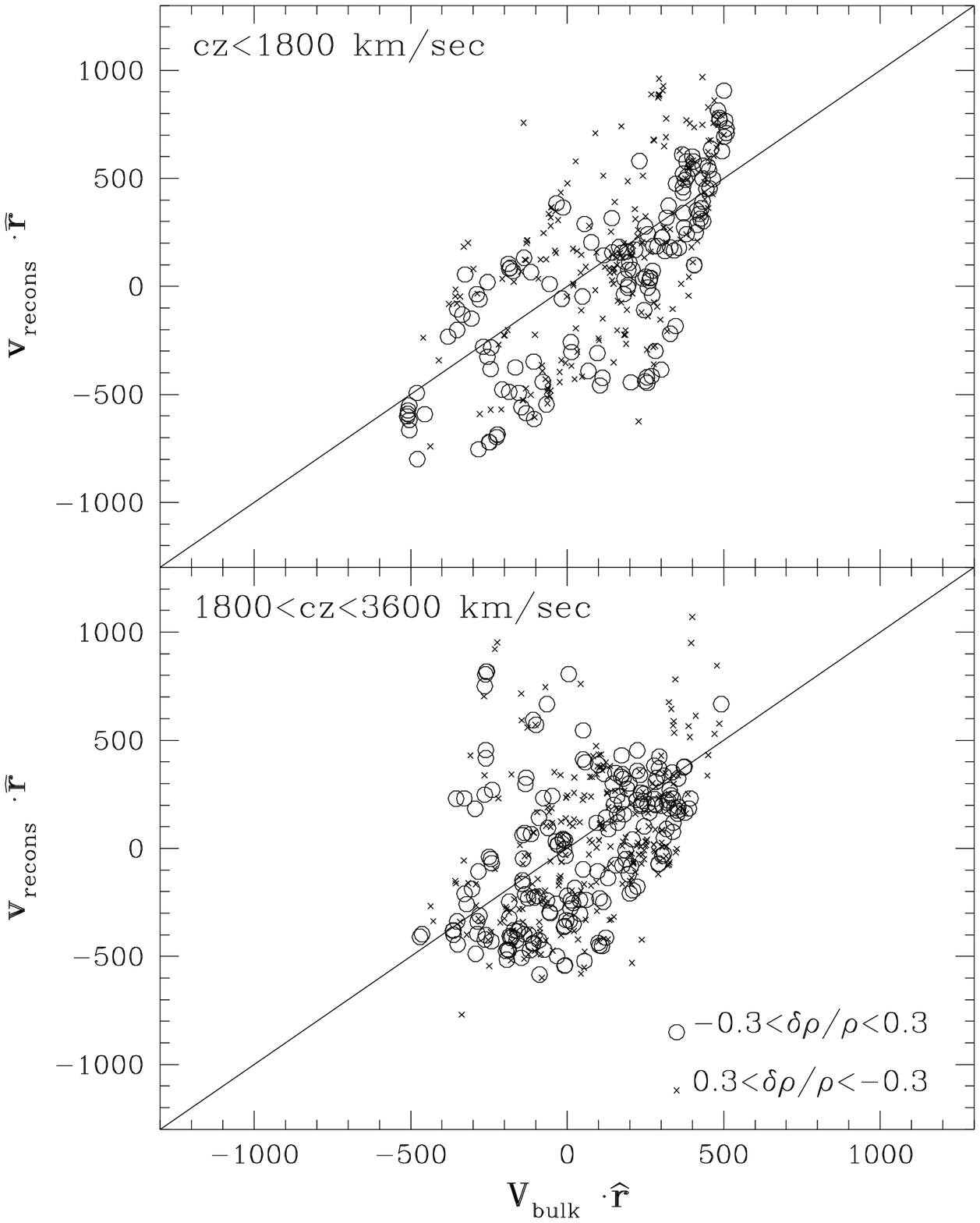}}
\caption{Comparison of local 1.2 Jy IRAS galaxy peculiar velocities provided 
by our model and by the full iterative algorithm within $cz \sim 3600$ km/sec
(see text). The diagonal line corresponds to ${\bf v}_{recons} \cdot 
\hat{\bf r} = {\bf V}_{bulk} \cdot \hat{\bf r}$.}
\end{figure}

However, many studies indicate that local galaxies have 
peculiar velocities not randomly oriented but rather participating in a 
coherent flow (bulk motion) together with the Local Group (ie., ${\bf v}(r)
\approx {\bf v}(0)$) within at least a volume of radius $\sim 5000$ km/sec 
(cf. Lynden-Bell et al 1988, Dekel 1994; 1997, Strauss \& Willick 1995). 
If so, it would be reasonable to 
evaluate the IRAS dipole in the LG frame, since in this case 
$cz \approx H_{\circ} r$. However, this is not absolutely true since there 
should exist also a velocity component due to the local, non-linear, dynamics 
acting between nearby galaxies and/or clusters of galaxies. We can therefore 
view the galaxy peculiar velocities as consisting of two vector components; 
a bulk flow and a local non-linear term:
\be\label{eq:v-mod}
{\bf v}(r)={\bf V}_{bulk}(r)+{\bf v}_{nl}(r)
\ee
Inserting eq.(\ref{eq:v-mod}) in eq.(\ref{eq:hubble}) and assuming that
${\bf v}(r) \cdot \hat{\bf r} \approx {\bf V}_{bulk}(r) \cdot \hat{\bf r}$ , 
ie., that the dominant component is that of the bulk flow \footnote{in a 
sense we assume that the vector average of ${\bf v}_{nl}(r) \cdot \hat{\bf r}$ 
over a whole sky distribution of galaxies is $\simeq 0$; not an unreasonable 
assumption in the limit of dense sampling.} 
we can use the observed bulk flow profile,
as a function of distance, given by Dekel (1994; 1997) and combined with 
that of Branchini, Plionis \& Sciama (1996) to correct the galaxy redshifts.
The zero-point, $V_{bulk}(0)$, and the direction of the bulk flow is estimated
applying eq.(\ref{eq:v-mod}) at $r=0$ and assuming, due to the ``coldness'' 
of the local velocity field (cf. Peebles 1988), that ${\bf v}_{nl}(0) \simeq 
{\bf v}_{inf}$ (where 
$v_{inf}$ is the LG infall velocity to the Virgo Supercluster). Using the
average value from the literature, ie. $v_{inf} \simeq 170$ km/sec, we obtain
$V_{bulk}(0) \simeq 500$ km/sec towards $(l,b) \simeq (276^{\circ},
15^{\circ})$.

We test our model by comparing peculiar velocities that it provides with 
those resulting from the full dynamical algorithm (kindly provided by Dr. 
Enzo Branchini) which estimates, using linear theory, the gravitational
acceleration at the position of each galaxy and then recovers the real-space 
galaxy distances by solving iteratively the generalised Hubble law of 
eq.\ref{eq:hubble} (cf. Yahil et al 1991; Strauss et al 1992).
In figure 3 we present this comparison
for relatively local galaxies in regions of $\delta\rho/\rho < 1$ (since at
dense regions the non-linear component that we neglect in our model will
dominate the galaxy peculiar velocity). 
We find a good correlation within $cz\sim 4000$ 
km/sec which is in fact the region where such corrections can affect the 
dipole. The correlation, at larger distances, progressively fades away since 
the bulk flow amplitude is low at such distances and the galaxy peculiar 
velocities are dominated by the distant {\em local} dynamics. In any case at 
such distances we do have $\int {\bf v}(r) \cdot \hat{\bf r} \; \rm{d}^{3}r 
\approx 0$ and thus redshift space distortions are dominated by the LG term 
in eq.(\ref{eq:hubble}) for which we do indeed correct the galaxy redshifts.
Note that we have verified that the amount of scatter seen in figure 3 is 
well reproduced from our model if we include a randomly oriented non-linear
velocity component with $\langle ({\bf v}_{nl} \cdot \hat{\bf r})^{2} 
\rangle^{1/2} \approx 320$ km/sec. 

We have further tested the robustness of the recovered real-space 
distribution by performing 200 Monte-Carlo simulations in which we vary 
$v_{inf}$ (and therefore also the amplitude and slightly the direction of 
${\bf V}_{bulk}(0)$) as well 
as the amplitude of $V_{bulk}(r)$ for all $r$'s, by randomly
sampling a Gaussian having as mean ($\mu$) the nominal velocity values and 
$\sigma=2\mu/3$. Furthermore, we investigate how our results change
when using the bulk-flow direction of Lauer \& Postman (1994); ie.,
$(l,b) \simeq (343^{\circ},52^{\circ})$ with $|V_{bulk}(r)| = 650$ km/sec
for $r\le 130$ $h^{-1}$ Mpc.

Finally, we would like to point out that it so happens that the IRAS dipole,
estimated in either the LG or the CMB frames, which should provide a sort of 
upper and lower dipole bounds respectively, differs very little and thus 
the $z$ to $3d$ frame correction {\em does not have a major consequence in our
main dipole results}. We do however investigate, in section 4.2, possible 
systematic effects that could be introduced by the frame transformation 
procedure in our IRAS-CMB dipole alignment results. 

\section{Main Results}
In figure 4a we present the two IRAS dipoles in redshift space.
We observe that they are consistent, although up to $\sim 50 \; 
h^{-1}$ Mpc the 1.2 Jy dipole is systematically higher than the QDOT dipole.
However, once we correct for redshift space distortions (figure 4b) the two 
corrected dipoles almost coincide within 50 $h^{-1}$ Mpc, as they should 
since both samples, due to their relative low flux limits, are good tracers 
of the Local Universe. 
Up to $\sim 130 \; h^{-1}$ Mpc the amplitudes of the two real-space dipoles 
deviate with the QDOT being larger than the IRAS 1.2Jy, which is to be expected
since the QDOT sample has a lower flux limit and thus it can `see' the
distant matter fluctuations better than the 1.2 Jy sample.
Beyond $\sim 130 \; h^{-1}$ Mpc, however, the two IRAS dipoles coincide again.
\begin{figure}
\mbox{\epsfxsize=9 cm \epsfysize=12cm \epsffile{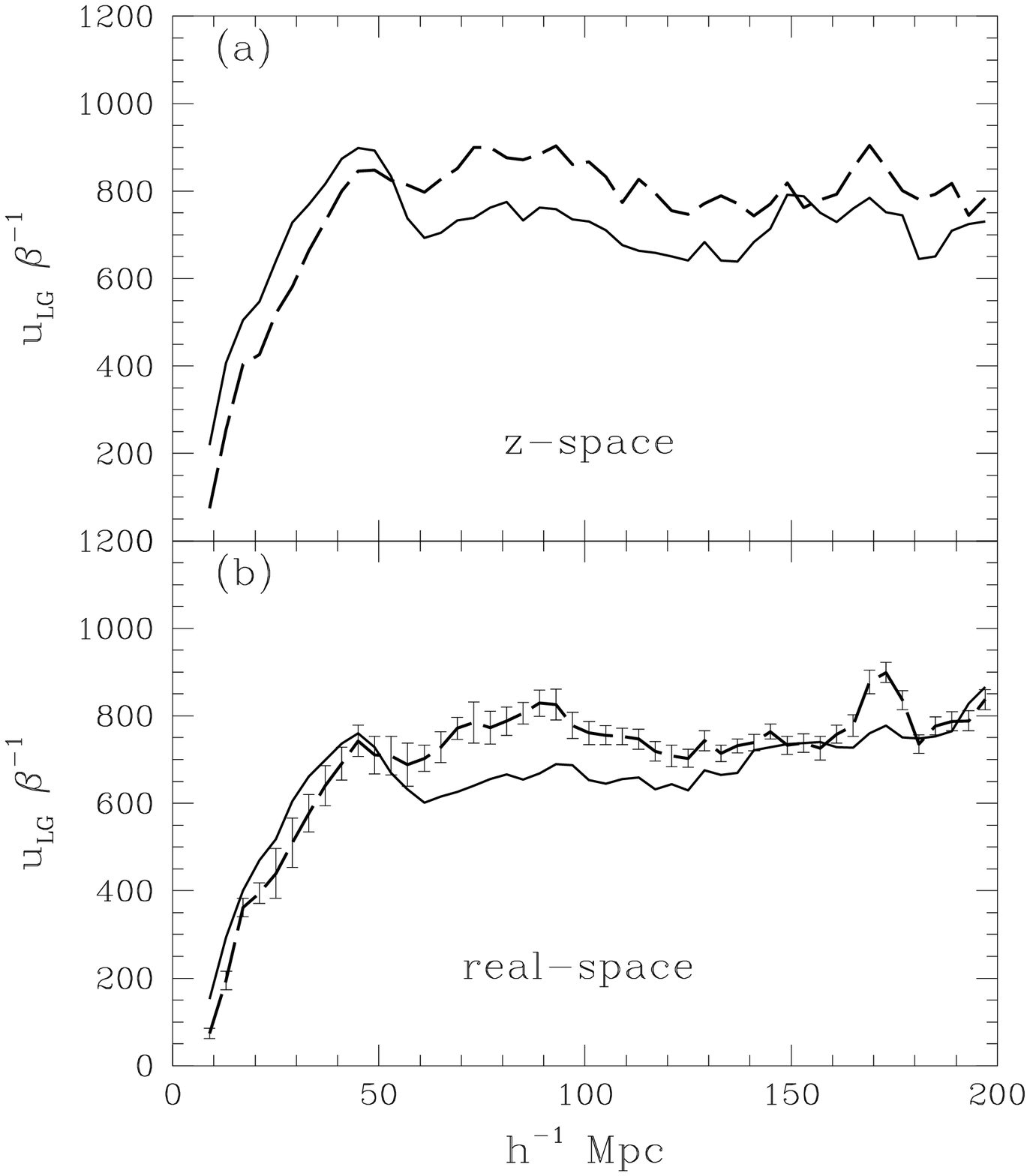}}
\caption{(a) IRAS 1.2 Jy (continuous line) and QDOT (dashed line) dipoles in 
redshift-space. (b) the corresponding
dipoles in real-space. 
The errorbars (shown only for the QDOT dipole) are 
estimated from Monte-Carlo simulations of the velocity field model (see text).}
\end{figure}
As can be seen from figure 4, redshift space distortions enhance significantly
the real-space dipole (by $\sim$ 12\%) only within $\sim 50$ $h^{-1}$ Mpc.
The uncertainties of the velocity field model, probed by
the Monte-Carlo simulations discussed previously, introduce a small scatter 
in the real-space dipole as indicated by the errorbars in figure 3b.
Using the Lauer \& Postman (1994) bulk-flow leaves unaltered
our main dipole results with only a small ($\sim 40$ km/sec) amplitude 
decrease.

\subsection{Evidence for $> 100$ $h^{-1}$ Mpc dipole contributions}
Between $\sim 150-180 \; h^{-1}$ Mpc there is an apparent amplitude bump, 
seen in both redshift and real-space IRAS dipoles. This bump is accompanied by 
a $\sim 5^{\circ}$ decrease of the misalignment angle between the two IRAS 
dipoles and that of the CMB, the 
overall misalignment angles at $r=200$ $h^{-1}$ Mpc being 
$\sim 23^{\circ}$ and $\sim 35^{\circ}$ for the 1.2 Jy and QDOT samples,
respectively. These facts suggest that this dipole amplitude bump is
not due to shot-noise uncertainties but rather it is an intrinsic effect, 
indicating the existence of contributions to the Local Group motion
from such large scales. Such contributions cannot be accurately determined,
however, from the present flux-limited samples and deeper samples are 
required for such a task (see Kolokotronis et al 1996).
\begin{table}
\caption[]{Differential 1.2 Jy IRAS dipole directions, the corresponding 
misalignment angles with respect to the CMB dipole, the dipole signal to 
noise ratio and probabilities of alignment within $\delta\theta$ (see text 
for complete definition).
The errorbars are estimated from Monte-Carlo simulations of the velocity 
field model, with the last two shells having no corresponding uncertainties
because we take $V_{bulk}(r) = 0$ for $r>170 \; h^{-1}$ Mpc (see text).}
\tabcolsep 2pt
\begin{tabular}{cccccccc} 
$h^{-1}$Mpc &  $N_{gal}$ & $S/N$ & $l^{\circ}$ & $b^{\circ}$  & 
$\delta\theta_{cmb}$ & $p_{mc}$ & $p_f$ \\ \\
5-110 & 4329 & 4.7$\pm 0.2$ & 246.0 & 35.5 & 26$\pm 3$ & 0.003 & 0.051 \\
110-139 &  390 & 0.6$\pm 0.2$ &311.8 &  84.1 &  55$\pm 17$ & 0.522 & 0.213 \\
139-159 & 220 & 1.7$\pm 0.2$ & 268.9 &  8.5 & 22$\pm 4$  &  0.026 & 0.036 \\
159-175 & 128 & 0.6$\pm 0.2$ & 307.1 &  31.5 & 26$\pm 7$  & 0.073 & 0.051 \\
175-188 &  72 & 0.0 & 96.6 &  61.4 & 89   & 0.655  & 0.492 \\
188-200 & 59 & 1.6 & 282.8 &  43.2 &  14   & 0.029 & 0.016
\end{tabular}
\end{table}
To further investigate these probable deep IRAS dipole contributions we
estimate the differential dipole in large equal volume shells (see 
Plionis, Coles \& Catelan 1993 for an earlier attempt in $z$-space). We
investigate shell sizes ranging from $2.5 \times 10^{6}$ to $8.2 \times
10^{6} \; h^{-3}$ Mpc$^{3}$ which give qualitatively similar results.
\begin{table}
\caption[]{QDOT dipole results (as in Table 1).}
\tabcolsep 2pt
\begin{tabular}{cccccccc} 
$h^{-1}$Mpc &  $N_{gal}$ & $S/N$ & $l^{\circ}$ & $b^{\circ}$  & 
$\delta\theta_{cmb}$ & $p_{mc}$ & $p_f$ \\ \\
5-110 & 1267 & 2.3$\pm 0.2$ & 232.5 & 30.3 & 38$\pm 3$ & 0.016 & 0.011 \\
110-139 & 210 & -0.1$\pm 0.2$ & 98.8 & 36.6 & 113$\pm 26$ & 0.89 & 0.71  \\
139-159 & 96 & 1.2$\pm 0.2$ & 271.1 & -12.7 &  43$\pm 9$ & 0.116 & 0.134 \\
159-175 & 66 & 1.2$\pm 0.1$ & 272.1 & 31.6 & 5$\pm 2$ & 0.004 & 0.002 \\
175-188 & 53 & -0.9 & 164.4 & -73.7 & 125 & 0.55 & 0.78 \\
188-200 & 31 & 0.7 & 318.4 & -15.6 & 60 & 0.55 & 0.25 
\end{tabular}
\end{table}
In tables 1 \& 2 we present the differential dipole directions and misalignment
angles with respect to the CMB dipole as well as a 
measure of the significance of the dipole of each individual shell, given by:
\be 
\frac{S}{N}=\frac{D_{raw}}{\sigma_{3D}} \cos(\delta\theta_{cmb}) \;.
\ee
for the case of $\delta V \simeq 5.5 \times 10^{6} \; h^{-3}$ Mpc$^{3}$
(six shells).
We observe that for the QDOT sample there are 3 shells with relatively small 
misalignment angles and dipole signal to noise ratios $>1$, the deepest 
shell being $[159 - 175] \; h^{-1}$ Mpc, in which $\delta\theta_{cmb} \sim 
5^{\circ}$. For the 1.2 Jy sample, which although shallower has a better 
sampling, we have small misalignment angles ($\delta\theta_{cmb} \mincir 
27^{\circ}$) in the same shells but also in a deeper shell ($188 - 200 \; 
h^{-1}$ Mpc). Out of these four aligned shells there are significant dipole 
contributions ($S/N>1$) only in three while the probability that these 
alignments are random is extremely low. 
The formal probability that two vectors are aligned within $\delta\theta$
is given by the ratio of the solid angle which 
corresponds to $\delta\theta$, to the solid angle of the whole sphere, ie.,
$p_{f}(\delta\theta) = \sin^{2}(\delta\theta/2)$.

We can now estimate the joint probability of alignment, 
within the observed $\delta\theta_{cmb}$, of $N$ independent vectors, 
which is given by:
\be
P^{N} \approx \prod_{i=1}^{N} p_{i}(\delta\theta)/p_{i}(90^{\circ})
\ee
Between three shells (first, fourth and sixth) the IRAS galaxy correlation 
function is zero, due to the large distances involved, and consequently the 
shells can be considered independent. 
Due, however, to the vicinity and therefore the possible 
correlation between the third and fourth QDOT shell we will consider their
joint probabilities as limits. 
Therefore, we have that the joint probability of alignment between the 
CMB and the differential IRAS equal-volume dipole directions (for those
with significant dipole signal $S/N>1$) is:
$$2 \times 10^{-4} \mincir P^{2,3}_{\rm QDOT} \mincir 8 \times 10^{-4}$$
$$P^{3}_{\rm 1.2Jy} \simeq 2 \times 10^{-4} \;.$$
for the QDOT and IRAS 1.2 Jy samples, respectively.

\subsection{Test for systematic alignment errors}
The observed differential dipole alignments could in principle be due to 
errors in 
the correction used to recover the 3-d frame in which we measure the dipole. 
For example, if redshift errors especially at large 
distances, were the sampling rate is low, were comparable to a significant 
fraction of the LG velocity, then using the 3-d galaxy distances estimated 
from eqs (\ref{eq:hubble}), (\ref{eq:v-mod}) and $v(0)=622$ km/sec, we could 
artificially produce a false alignment of the distant shells differential 
dipole with that of the CMB.
We therefore address the question of which are the dipole alignments induced 
totally due to our frame transformation procedure ie., for the extreme case
where there is no intrinsic dipole and no LG peculiar velocity. 
We run 5000 Monte-Carlo simulations in 
which we destroy the intrinsic IRAS galaxy dipole as well as redshift space 
distortions by randomising the angular coordinates of the galaxies while 
keeping their distances and therefore their selection function unchanged. 
On this intrinsically random galaxy distribution we apply our $z$ to 
$3d$ space transformation and then measure the artificially induced 
differential dipole alignments due to the frame transformation itself. 
The coupling between the space distortion and the selection 
function results in a non-trivial alignment behaviour. In the first shell we 
observe anti-alignments while at more distant shells the artificial 
alignment effect does appear. For example at the last shell the median 
$\delta\theta$ is $\sim 67^{\circ}$ instead of 90$^{\circ}$. However, it 
is impossible to create the observed IRAS dipole alignments if 
there is no true signal present.
We quantify this by measuring the probability, $p_{mc}(\delta\theta)$, of 
observing in our Monte-Carlo simulations dipole alignments as large as
the observed IRAS differential dipole alignments. For the shells of interest
we find that this probability is low and comparable to the expected 
$p_{f}$, which implies that 
the frame transformation uncertainties cannot induce the observed IRAS dipole 
alignments (see the corresponding values of $p_{mc}$ and $p_f$ in the tables).

We conclude that the differential dipole directions are not randomly oriented
with respect to the CMB and therefore we do not only have indications 
for significant dipole contributions from large depths but also for a coherent 
anisotropy extending to these large scales.

\subsection{Possible cause of the large-scale IRAS dipole contributions}
It is interesting that stronge evidence exist for deep 
dipole contributions from the available galaxy cluster data. 
Contributions up to $\sim 20\% - 30\%$ of the total cumulative optical 
and X-ray cluster dipole, were found from $\sim 140 - 
160$ $h^{-1}$ Mpc depths (Scaramella et al 1991; Plionis \& Valdarnini 1991; 
Branchini \& Plionis 1996; Plionis \& Kolokotronis 1998). 
Similar {\em coherence} of the differential dipoles in equal volume shells
was also found in galaxy cluster case (Plionis \& Valdarnini 1991; Plionis et 
al 1993). These studies have shown that the cause of the deep dipole 
contributions should be 
attributed mostly to the Shapley concentration, a huge mass overdensity 
located at $\sim 140$ $h^{-1}$ Mpc in the general direction of the 
Hydra-Centaurus supercluster (Shapley 1930; Scaramella et al 1989; 
Raychaudhury 1989).

To investigate further the possible cause of the present IRAS dipole results
we have smoothed the IRAS 1.2 Jy galaxy distribution in a $40^{3}$ cube with
a cell size of 10 $h^{-1}$ Mpc using a Gaussian with smoothing radius
equal to one cell and weighting each galaxy by $\phi^{-1}$. 
Due to the coupling 
between the selection function and the constant radius smoothing, we correct
the resulting smoothed distribution for a distance dependent effect, which we 
quantified using N-body simulations (details will
be presented in a forthcoming paper). In figure 5 we present the smoothed
IRAS 1.2 Jy galaxy distribution on the supergalactic plane (of 10 $h^{-1}$ 
Mpc width) within 170 $h^{-1}$ Mpc. The contour step is 0.4 in 
overdensity while the $\delta\rho/\rho=0$ level appears as a thick continuous 
line. 

Well known structures appear in this plot; the largest and most evident
is the Shapley concentration located at $(X_{sup},Y_{sup})\approx (-120, 60)$,
the Perseus-Pisces supercluster at $(X_{sup},Y_{sup})\approx (60, -40)$, the 
Coma supercluster at $(X_{sup},Y_{sup})\approx (-20, 70)$, the Ursa-Major
supercluster at $(X_{sup},Y_{sup})\approx (100, 100)$, the Pisces-Cetus
supercluster at $(X_{sup},Y_{sup})\approx (50, -140)$ while
the Great Attractor (Hydra-Centaurus complex ?), at $(X_{sup},Y_{sup})
\approx (-30, 30)$ appears in the foreground of the Shapley concentration. 
\begin{figure}
\mbox{\epsfxsize=8 cm\epsffile{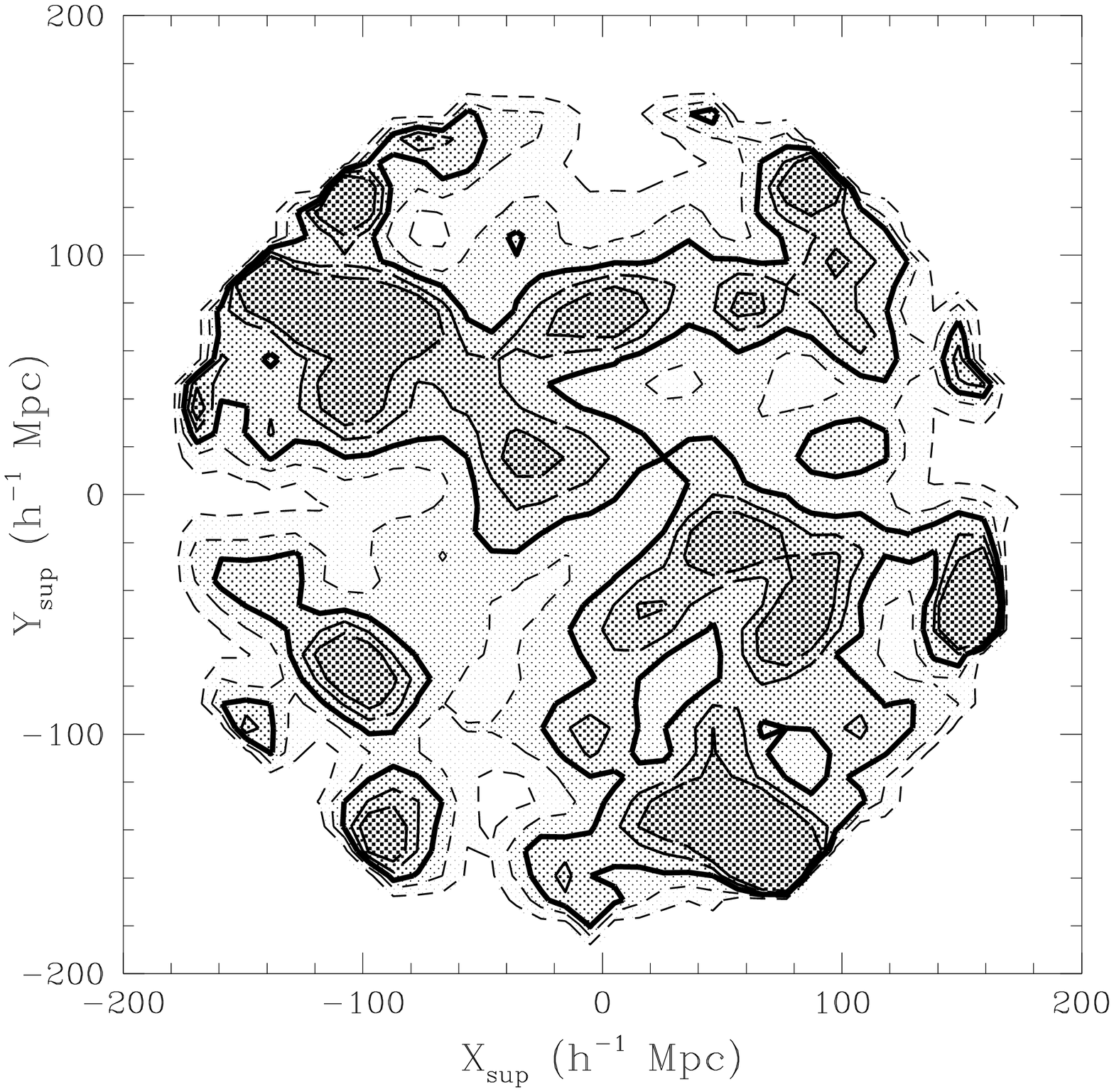}}
\caption[]{Contour plot of the smoothed 1.2 Jy IRAS galaxy 
distribution in real space and on the supergalactic plane.}
\end{figure}
Furthermore, figure 5 is very similar to the corresponding Abell/ACO cluster 
density field (cf. Tully et al 1992; Branchini \& Plionis 1996, their figure 
3), and we therefore obtain a consistent picture, from both IRAS 
galaxy and Abell/ACO cluster data, in which the Shapley concentration is the 
most probable cause of the deep dipole contributions while the general 
alignment of the Great Attractor, Perseus-Pisces and Shapley superclusters 
is most probably the cause of the apparent {\em coherence} of the IRAS galaxy 
dipole.

\subsection{$\Omega_{\circ}^{0.6}/b_{I}$ from the IRAS dipoles}
Using the real-space dipole results and equation (\ref{eq:lpt}), we can 
estimate the density parameter $\beta_{IRAS}$. 
However, the value obtained should be 
considered rather as an upper limit since the deep contributions to the dipole,
for which we do have strong indications, are probably not fully revealed 
by the present samples (see Kolokotronis et al 1996). 
Taking into account the scatter among the two IRAS samples, the 
amplitude variations at large depths and the uncertainties of the
velocity model used to recover the real-space galaxy distances, we find:
\be 
\beta_{IRAS} \mincir 0.78 (\pm 0.1) \;,
\ee
in agreement with the QDOT analysis of Rowan-Robinson et al (1990) but 
slightly larger, although within 1$\sigma$,
than the 1.2 Jy results of Strauss et al (1992).
This value of $\beta_{I}$ implies either that $\Omega_{\circ}\mincir 0.66$ 
for $b_{IRAS}=1$ or $\Omega_{\circ}\simeq 1$ for $b_{IRAS} \magcir 1.28 $.

\section{Conclusions}
Using a consistent analysis procedure we find that within $50 \; h^{-1}$ Mpc, 
both the 1.2Jy and 0.6Jy (QDOT) IRAS samples, give identical dipole results
while beyond this depth the QDOT dipole increases substantially up to
$100 \; h^{-1}$ Mpc,  in agreement with Rowan-Robinson et al (1990). 
Furthermore there are significant indications for (a) dipole contributions 
from depths $\sim 170$ $h^{-1}$ Mpc, in agreement with other recent 
large-scale studies (cf. Plionis \& Kolokotronis 1998) and (b)
a coherence of the dipole anisotropy extending to similar depths. The most
probable cause of these deep dipole contributions is the Shapley mass
concentration, while of the dipole {\em coherence} is the general alignment, on
the supergalactic plane, of the Perseus-Pisces supercluster, the Great 
Attractor and the Shapley concentration, which span a range of $\sim 200 \; 
h^{-1}$ Mpc.

A similar study of the complete (6 in 6) IRAS 0.6 Jy sample (PSCz) should give 
better indications of these results, although the overall amplitude of the 
effect could be probably estimated by a deeper catalogue (limited at a 
lower flux limit).

\section* {Acknowledgements}
We thank Enzo Branchini for providing us with the reconstructed IRAS 
1.2 Jy galaxy peculiar velocities.

\end{document}